\documentclass{article}
\usepackage{spconf,amsmath,graphicx}
\usepackage{color}
\usepackage {booktabs}
\usepackage{amssymb}
\usepackage{comment}
\usepackage{mathrsfs}
\usepackage{amsmath}
\usepackage[
backend=biber,
style=ieee,
citestyle=numeric-comp,
maxbibnames=3,
maxcitenames=3,
doi=false,isbn=false,url=false,eprint=false
]{biblatex}

\usepackage{multirow}
\defbibheading{bibliography}[\refname]{}

\DeclareMathOperator*{\argmax}{arg\,max}

\title{HuBERTopic: Enhancing semantic representation of HuBERT \\through self-supervision utilizing topic model}
\name{Takashi Maekaku$^1$, Jiatong Shi$^2$, Xuankai Chang$^2$, Yuya Fujita$^1$, Shinji Watanabe$^2$}
\address{
  $^1$LY Corporation, Tokyo, JAPAN\\
  $^2$Carnegie Mellon University, PA, USA
}
\begin{document}
\ninept
\maketitle
\begin{abstract}
Recently, the usefulness of self-supervised representation learning (SSRL) methods has been confirmed in various downstream tasks. Many of these models, as exemplified by HuBERT and WavLM, use pseudo-labels generated from spectral features or the model's own representation features. 
From previous studies, it is known that the pseudo-labels contain semantic information.
However, the masked prediction task, the learning criterion of HuBERT, focuses on local contextual information and may not make effective use of global semantic information such as speaker, theme of speech, and so on.
In this paper, we propose a new approach to enrich the semantic representation of HuBERT.
We apply topic model to pseudo-labels to generate a topic label for each utterance. An auxiliary topic classification task is added to HuBERT by using topic labels as teachers.
This allows additional global semantic information to be incorporated in an unsupervised manner.
Experimental results demonstrate that our method achieves comparable or better performance than the baseline in most tasks, including automatic speech recognition and five out of the eight SUPERB tasks.
Moreover, we find that topic labels include various information about utterance, such as gender, speaker, and its theme.
This highlights the effectiveness of our approach in capturing multifaceted semantic nuances.

\end{abstract}
\section{Introduction}
\label{sec:intro}
\vspace{-8pt}
In recent years, the efficacy of self-supervised representation learning (SSRL) methodologies has been empirically validated across diverse downstream tasks in the speech community \cite{yang2021superb, tsai2022superb, xuankai2021asru, peng2023study, evain21_interspeech, shi23g_interspeech, conneau22_interspeech,javed2023indicsuperb,  mohamed2022self}. Prominent exemplars of these models, such as HuBERT \cite{hsu2020hubert}, Wav2Vec2.0 \cite{baevski2020wav2vec}, WavLM \cite{chen2022wavlm}, w2v-bert \cite{chung2021w2v}, and BEST-RQ \cite{chiu2022self}, predominantly rely on pseudo-labels generated from either mel-frequency cepstrum coefficients (MFCC) or the model's own representation features.
HuBERT, owing to its utilization of semantic contextual information both in its learning criterion through pseudo-units and in its modeling via self-attention networks stands out as a strong self-supervised learning (SSL) mechanism. It is worth noting that semantic information plays a pivotal role in a variety of tasks, including automatic speech recognition (ASR), spoken language understanding (SLU) \cite{istaiteh2023transformer}, and more.

In our previous work \cite{maekaku2023fully}, we demonstrated that we could effectively cluster a set of conversational speech based on the theme of the conversation by applying the topic model to pseudo-labels from HuBERT as input.
This highlighted that the pseudo-labels inherently contain some semantic information, and
this fact is also supported by the studies of \cite{polyak2021speech} and \cite{wu2023improving}.
However, HuBERT's masked prediction task primarily focuses on the information surrounding the masked regions. Consequently, even if the discrete sequences inherently capture semantic information, they may not effectively harness the insights derived from the broader global context.
In other words, there is the possibility that the performance of different tasks can be expected to improve by feeding some global semantic information explicitly back to HuBERT.

In this paper, we propose HuBERTopic, which is a novel approach to enrich the semantic representation of HuBERT.
We apply the topic model to pseudo-labels to generate a topic label for each utterance. 
To capture the global semantic information of the entire speech, a CLS vector is given at the beginning of the input to the Transformer encoder of HuBERT, and an auxiliary topic classification task is added by using topic labels as teachers.
This allows additional global semantic information that is expected to be useful in improving the performance of downstream tasks to be incorporated in an unsupervised manner.

There have been several efforts to improve the performance of downstream tasks by applying additional unsupervised methods to HuBERT. ContentVec \cite{qian2022contentvec} and Spin \cite{chang2023self} are self-supervised methods that disentangle speaker information to impose a speaker-invariant constraint on a pre-trained HuBERT. Shi \textit{et al.} utilized unsupervised ASR to bridge speech SSL with text-pre-trained models in order to take benefits from semantic information from the textual representations \cite{shi2023bridging}.
These methods have been successful in a variety of downstream tasks, but our method differs in that HuBERTopic does not focus on specific attributes such as speaker and phoneme, and it trains to enrich the global semantic information during pre-training.
Moreover, in the field of SSL, methodologies leveraging some information obtained from the topic model as the ground truth labels have not been previously explored. Within the domain of Natural Language Processing (NLP), various approaches have been proposed that enhance the topic estimation of Latent Dirichlet Allocation (LDA) \cite{blei2003latent} through a semantic filter based on BERT \cite{kenton2019bert} \cite{venugopalan2022enhanced}, or attempt to estimate topic labels using BERT and clustering technique \cite{grootendorst2022bertopic}. However, it is worth noting that, to the best of our knowledge, the utilization of topic information as explicit teacher labels has not been proposed to date.

Our experimental results demonstrate that our method outperformed the baseline performance on most tasks, including ASR task and eight SUPERB tasks.
Moreover, we find that pseudo-labels include various information about the utterance, such as gender and speaker, through a topic model.
This highlights the effectiveness of our approach in capturing multifaceted semantic nuances, ultimately leading to improved performance in downstream tasks.

\vspace{-10pt}
\section{Methodology}
\begin{figure}[tbp]
\centering
  \includegraphics[clip,width=8.5cm]{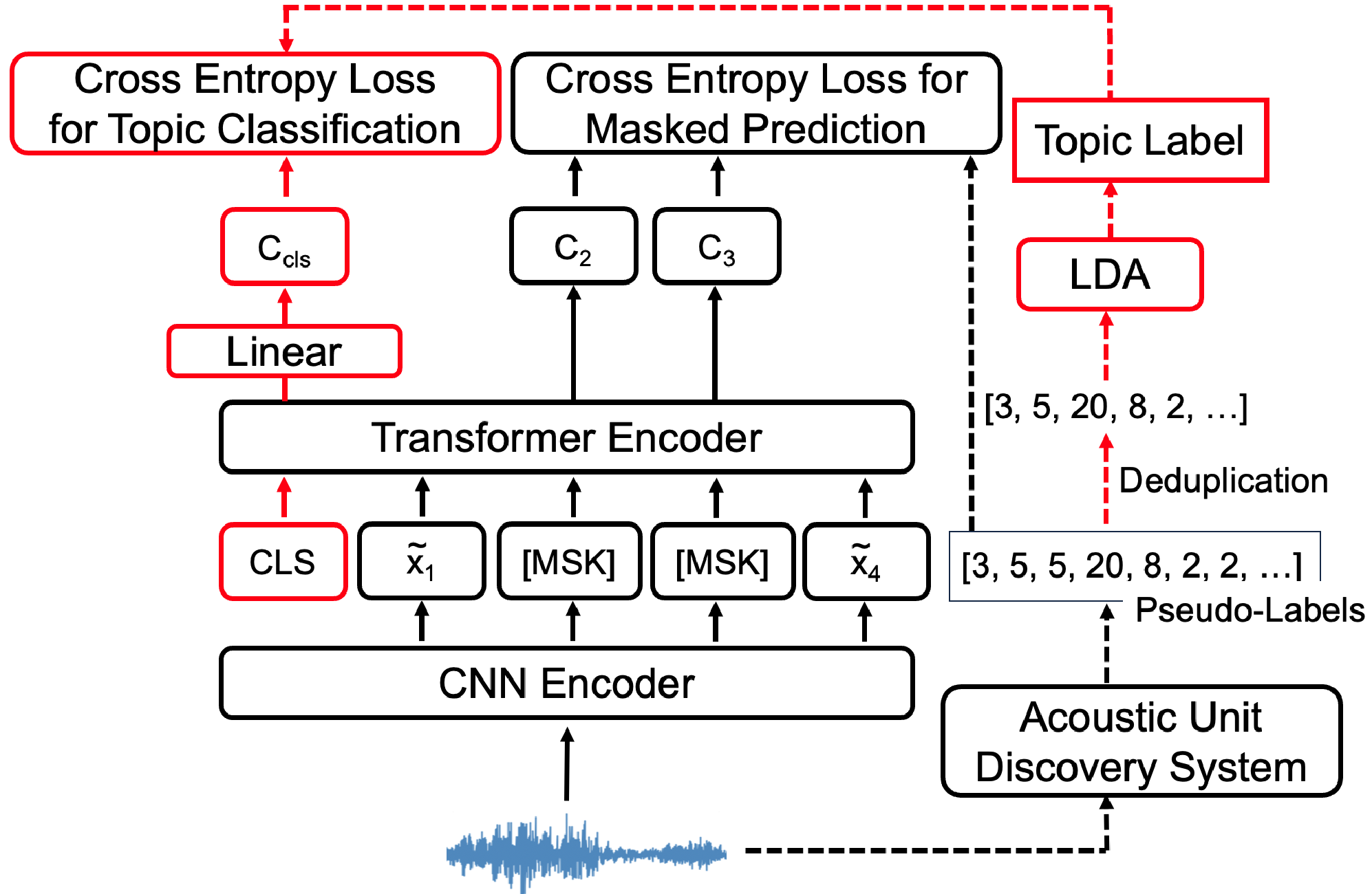}
  \vspace{-10pt}
  \caption{\textrm HuBERTopic network structure. The areas in black are the original HuBERT network and the areas in red are the parts extended by the proposed method.}
  \label{fig:hubertopic}
  \vspace{-15pt}
\end{figure}

\vspace{-5pt}
\label{sec:proposed}
This section describes the proposed method, HuBERTopic, and the techniques used in the method.
The proposed network structure is shown in Figure~\ref{fig:hubertopic}.
The areas in black are the original HuBERT network, and the areas in red are the parts extended by the proposed method.
In the proposed method, a topic label is generated for each utterance by LDA, using the de-duplicated pseudo-label sequence for each utterance for HuBERT training as pseudo-text.
The labels are then used to impose a topic label classification task on HuBERT.
This makes it possible to provide the model with richer global semantic information about the utterances whose topic labels contain in an unsupervised manner.
The details of each method are described below.

\vspace{-5pt}
\subsection{HuBERT}
\label{subsec:hubert}

HuBERT \cite{hsu2020hubert} is a self-supervised representation learning method based on a masked prediction of pseudo-discrete labels.
The masked prediction model consists of a convolutional neural network (CNN) feature encoder and a Transformer encoder.
HuBERT refines pseudo-labels using an iterative training framework.
In the first iteration, the pseudo-labels are generated by applying k-means clustering to the MFCC features of input speech.
We define a set of raw waveform as $\mathcal{X}=\{X^d|d=1,\cdots,D\}$, where $X^d=(x_t^d|t=1,\cdots,T^d)$. $D$ is the total number of utterances and $T^d$ is and the length of $d$-th utterance.

Let us denote pseudo-labels obtained by k-means clustering of MFCC features as $Z^d=(z_t^d \in \{1,\cdots,U\}|t=1,\cdots,M^d)$, where $U$ is the number of cluster units of k-means clustering and $M^d$ the length of $d$-th label sequence.
Then, by denoting a set of indices to be masked and the masked acoustic feature sequence as $\mathcal{M}^d \subset \{1,\cdots,M^d\}$ and $\tilde{X}^d$, respectively,
the pre-training objective function is described as follows:
\vspace{-5pt}
\begin{equation}
\label{hubert_objective}
    \mathcal{L}_{\text{MP}}(\mathcal{X})=-\sum _{d} \sum_{t\in \mathcal{M}^d}\log p_f(z_t^d|\tilde{X}^d),
\end{equation}
where $p_f(\cdot)$ is the masked prediction model.
For further improvement of performance, the pseudo labels are re-generated from the second stage of the pre-training using an intermediate representation of the masked prediction model pre-trained at the previous stage.
In this study, we consider the pseudo-labels $Z^d$ with the same adjacent labels deduplicated as pseudo-text $Z'^d$, and apply a topic model described below.
\vspace{-10pt}

\subsection{Topic Label Generation using Latent Dirichlet Allocation}
\vspace{-5pt}
\label{lda}
This study uses Latent Dirichlet Allocation (LDA) \cite{blei2003latent} as the topic model.
LDA is a probabilistic generative model for estimating latent topics inherent in discrete data such as text.
Here, we consider the pseudo-text obtained in Section~\ref{subsec:hubert} as a single document and estimate the topic label that this text has.
Let us denote a pseudo-text corpus as
$\mathcal{Z^{\prime}} = \{Z'^1, \cdots, Z'^D\}$
where $Z'^d = ({z^{\prime}}_m^d|m=1,\cdots,M'^d)$ and $M'^d$ is the length of $d$-th pseudo-text. ${z^{\prime}}_m^d$ is the pseudo-label that is an instance from a set of pseudo-label vocabulary $\mathcal{C}$.

LDA assumes a text can be represented by a mixture of probability distributions over words depending on the latent topic. In this paper, we replace the text with the pseudo-text and words with the pseudo-label, respectively.
It follows the generative model described below:
\begin{itemize}
\setlength{\itemsep}{-0.7mm}
\setlength{\parskip}{-0.9mm}
    \item[] For each pseudo-text in a corpus,
    \begin{enumerate}
    \setlength{\itemsep}{-0.6mm}
    \setlength{\parskip}{0.0mm}
        \item Draw a length of pseudo-text $M'^d$ from a Poisson distribution.
        \item Draw $\theta^{d}$ from a Dirichlet distribution $\text{Dir}(\alpha)$.
        \item For each pseudo-label ${z^{\prime}}_m^d$ in the pseudo-text,
        \begin{enumerate}
        \setlength{\itemsep}{-0.5mm}
            \setlength{\parskip}{0.0mm}
            \item Draw a latent topic $s^d_m$ from Multinomial distribution $\text{Multinomial}(\theta^d)$ whose parameter is sampled at the previous step.
            \item Draw a pseudo-label from probability distributions over words $p({z^{\prime}}_m^d |s^d_m, \beta)$ corresponding to the latent topic $s^d_m$ sampled at the previous step.
        \end{enumerate}
    \end{enumerate}
\end{itemize}
Here $\theta^d\in (0,1)^K$ is the topic proportion vector, which represents what distribution of topics the $d$-th pseudo-text is composed of.
$K$ is the number of topics.
$s^d_m \in \{1,\cdots,K\}$ is the latent topic of $m$-th pseudo-label of pseudo-text $Z'^d$. 
$\alpha$ is the hyper-parameter of the Dirichlet distribution and $\beta$ is the probability distribution over pseudo-labels of all the latent topics.

A necessary part of inference using LDA is to compute the posterior distribution with respect to $\Theta = \{\theta^1, \cdots \theta^D\}$ and $\mathcal{S}=\{S^1, \cdots S^D\}$ where $S^d = \{s^d_1, \cdots s^d_{M'^d}\}$ given a set of pseudo-text $\mathcal{Z'}$ as follows:
\begin{align}
\label{posterior}
p(\Theta, \mathcal{S} | \mathcal{Z'},\boldsymbol{\alpha},\boldsymbol{\beta}) = \frac{p(\Theta, \mathcal{S}, \mathcal{Z'}|\boldsymbol{\alpha},\boldsymbol{\beta})}{p(\mathcal{Z'}|\boldsymbol{\alpha},\boldsymbol{\beta})}.
\end{align}
However, this distribution is intractable to compute in general.
In practice, an approximation of the posterior is obtained using variational inference \cite{blei2003latent}. It can be approximated as the product of the variational distributions $q(\Theta)$ and $q(\mathcal{S})$ on the left side of Eq.~\ref{posterior}, and the posterior distribution of $\theta^d$, $p(\theta^d|Z'^d)$ is approximated as $q(\theta^d)$.

In this study, we assign one topic label per pseudo-text, i.e., per utterance. Let this topic label for pseudo-text $Z'_d$ be a $\lambda_{Z'_d}$, this is obtained by
\vspace{-5pt}
\begin{align}
\label{eq:topic_label}
    \lambda_{Z'_d} = \argmax _{k} p(\theta^d|Z'^d) \approx \argmax _{k} q(\theta^d).
\end{align}
\begin{table*}[t]
    \centering
    \caption{ASR evaluation results. In the case of 1-iteration, the number of topics $K$ for 0-iteration and $K$ for 1-iteration are listed in order. Detailed settings are discussed in Sec.~\ref{sec:experimental_setup}.}
    \vspace{-5pt}
\begin{tabular}{cccccccc}
\toprule
\textbf{Iteration} & \textbf{Training Data} & \textbf{Model} & $K$ & \multicolumn{4}{c}{\textbf{WER($\downarrow$)}}   \\
& & & & \textbf{dev-clean} & \textbf{dev-other} & \textbf{test-clean} & \textbf{test-other}  \\
\midrule
\multirow{6}{*}{0} & \multirow{6}{*}{LS-100h} & HuBERT & - & 21.7 & 38.6 & 22.0 & 40.0  \\
& & \multirow{5}{*}{HuBERTopic} & 2 & 20.7 & 37.2 & 21.0 & 39.0  \\
& & & 10 & 21.2 & 37.9 & 21.2 & 39.7 \\
& & & 30 & \textbf{20.2} & \textbf{36.5} & \textbf{20.2} & \textbf{38.3} \\
& & & 60 & 20.4 & 37.3 & 20.8 & 39.0 \\
& & & 90 & 20.5 & 36.9 & 20.7 & 38.6 \\

\midrule
\multirow{6}{*}{1} & \multirow{6}{*}{LS-100h} & HuBERT & - & 17.1 & 33.5 & 17.3 & 35.3 \\
& & \multirow{5}{*}{HuBERTopic} & 30-30 & 16.2 & \textbf{32.7} & 16.7 & 34.2 \\
& & & 30-100 & 16.2 & 33.0 & 16.7 & 34.6 \\
& & & 30-200 & \textbf{16.1} & 32.9 & \textbf{16.6} & \textbf{34.1} \\
& & & 30-300 & 16.3 & 33.3 & 16.8 & 34.4 \\
& & & 30-400 & 16.2 & 32.9 & 16.6 & 34.4 \\

\midrule
\multirow{4}{*}{1} & \multirow{4}{*}{LS-960h} & HuBERT & - & 7.4 & 14.2 & \textbf{7.4} & 14.2 \\
& & \multirow{3}{*}{HuBERTopic} & 30-30 & \textbf{7.2} & \textbf{14.1} & \textbf{7.4} & \textbf{13.7} \\
& & & 150-1000 & 7.3 & 14.3 & 7.5 & 14.1 \\

\bottomrule
\end{tabular}
    \label{tab:asr}
    \vspace{-15pt}
\end{table*}

\vspace{-15pt}
\subsection{Imposing a Topic Classification Task on HuBERT}
HuBERTopic is structured with a topic label classification task added to HuBERT for the purpose of enhancing the semantic information in the representation features.
As shown in Figure~\ref{fig:hubertopic}, the CLS vector is added at the beginning of the input to the Transformer encoder to supplement the context of the entire utterance as in BERT \cite{kenton2019bert}.
The entire network is trained with multi-task learning.
Unlike BERT, the input to the encoder is a set of continuous vectors. Therefore, the CLS vector is a random vector of continuous values in this study.
After this vector is input to the Transformer encoder, it is fed through a linear layer, and the embedded representation $c_{\text{cls}}\in \mathbb{R}^K$ is output.
The loss function for topic classification $\mathcal{L}_{\text{TC}}$ is calculated 
as follows:
\begin{equation}
\mathcal{L}_{\text{TC}} = -\sum_{d}\sum_{k} \tau_k^d \log(\text{softmax}({c_{\text{cls}}}^d)_k),
\end{equation}
where $\tau^d$ is the one-hot representation of $\lambda_{Z'_d}$. It is a $K$-dimensional vector whose elements are 1 when $\lambda_{Z'_d}=k$ and 0 otherwise.
${c_{\text{cls}}}^d$ represents $c_{\text{cls}}$ for the $d$-th pseudo-text.
Finally, the total loss, denoted as $\mathcal{L}$, is calculated as a weighted sum of $\mathcal{L}_{\text{MP}}$ and $\mathcal{L}_{\text{TC}}$ as:

\vspace{-8pt}
\begin{align}
\label{multitask}
    \mathcal{L} = (1-\rho)\mathcal{L}_{\text{MP}} + \rho\mathcal{L}_{\text{TC}},
\end{align}
where $\rho$ is a mixing weight, determined through preliminary experiments, and was set to 0.01 in this work.
The masked prediction task learns local contextual information, but by subjecting HuBERT to the additional topic classification task, we expect it to simultaneously hold global semantic information that captures the context of the entire speech in an unsupervised manner.
Note that HuBERTopic learns without assuming the existence of specific attribute information such as speaker, as in \cite{qian2022contentvec,chang2023self}, etc. What information is contained in these topics is analyzed in Section~\ref{subsec:analysis}.

\vspace{-8pt}
\section{Experimental Setup}
\vspace{-8pt}
\label{sec:experimental_setup}

\textbf{Dataset and Tasks.}
All SSL models, including the baseline, were trained either on the full 960 hours of LibriSpeech (LS-960h) \cite{panayotov2015librispeech} or a 100-hr subset (LS-100h). For the ASR model, LS-100h was used for training, and \textit{dev-\{clean,other\}} and \textit{test-\{clean,other\}} were used for tuning and evaluation, respectively. 
Regarding the performance of the proposed method, we first tuned and evaluated the number of topics $K$ in the ASR task. 

The top-performing models from our experiments underwent further evaluation using the Speech Processing Universal PERformance Benchmark (SUPERB) framework \cite{yang2021superb, tsai2022superb}. This assessment adhered to the SUPERB guidelines for SSRL evaluation, employing a weighted sum of the \textit{frozen} hidden layers from the specified SSRL model. We conducted evaluations across a broad range of tasks, namely phoneme recognition (PR), emotion recognition (ER), intent classification (IC), speaker identification (SID), speaker diarization (SD), slot filling (SF), keyword spotting (KS), and speech enhancement (SE). These tasks comprehensively address the four core dimensions of speech: content, speaker characteristics, semantic meaning, and paralinguistic cues. For performance metrics, we followed the standards set by the SUPERB framework. Specifically, we used the phoneme error rate (PER) for PR,  diarization error rate (DER) for SD, F-1 measure for SF, perceptual evaluation of speech quality (PESQ) for SE, and accuracy as the metric for ER, IC, SID, and KS.
\begin{table*}[t]
    \centering
    \caption{SUPERB evaluation results, where the proposed HuBERTopic achieves comparable or better performances than baseline HuBERT over various SUPERB tasks. Detailed settings are discussed in Sec.~\ref{sec:experimental_setup}.}
    \vspace{-5pt}
\begin{tabular}{c|c|ccccccccccc}
\toprule
\textbf{Training Data} & \textbf{Model} & \textbf{PR($\downarrow$)} & \textbf{ER($\uparrow$)} & \textbf{IC($\uparrow$)} & \textbf{SID($\uparrow$)} & \textbf{SD($\downarrow$)} & \textbf{SF($\uparrow$)} & \textbf{KS($\uparrow$)} & \textbf{SE($\uparrow$)} \\
\midrule
\multirow{2}{*}{LS-100h} & HuBERT & 13.89 & 60.24 & 88.72 & 60.48 & 8.86 & 80.62 & 94.22 & 2.48 \\
& HuBERTopic & \textbf{12.97} & \textbf{60.92} & \textbf{90.64} & \textbf{61.82} & \textbf{8.59} & \textbf{81.05} & \textbf{94.87} & \textbf{2.50} \\
\midrule
\multirow{3}{*}{LS-960h} & HuBERT & 5.04  & \textbf{64.12} & 97.57 & \textbf{79.34} & 7.49 & 88.61 & \textbf{96.04} & 2.53 \\
& HuBERTopic (30-30) & \textbf{4.83} &  64.10 & 97.68 & 78.98 & \textbf{6.93} & 88.76 & 95.26 & 2.53 \\
& HuBERTopic (150-1000) & 4.84 &  63.61 & \textbf{98.10} & 79.21 & 7.07 & \textbf{88.79} & 95.81 & \textbf{2.55} \\
\bottomrule
\end{tabular}
    \label{tab:superb}
    \vspace{-15pt}
\end{table*}

\noindent
\textbf{Model Configuration.}
We utilized the identical architecture of HuBERT \texttt{BASE} \cite{hsu2020hubert} for the baseline model. 
For the proposed method, the CLS vector inserted as the first input to the Transformer encoder is a 512-dimensional random vector, which is not learnable.
Let $K'(\leq K)$ be the number of topic labels actually used in the classification task\footnote{Note that some topic labels may not correspond to any pseudo-text, since labels are selected by taking argmax from a $K$-dimensional vector for each pseudo-text, as in Eq.~\ref{eq:topic_label}. }, the linear layer inserted after the encoder prepared for the CLS vector is a $256\times K'$ weight matrix.
The number of iterative training for all models is 2. The first session uses the MFCC and the second session uses the 6th layer features of the HuBERT/HuBERTopic encoder to generate pseudo-label sequences by k-means.
Following the pre-training phase, we employ the connectionist temporal classification (CTC) \cite{graves2012connectionist} loss during the ASR fine-tuning process. The projection layer is eliminated and substituted with a randomly initialized softmax layer.
The implementation of the proposed method was based on ESPnet \cite{watanabe2018espnet}, and Gensim\footnote{https://radimrehurek.com/gensim/} was used to train LDA.
Eight V100 GPUs were used during HuBERT/HuBERTopic training with LS-960h, while eight A100 GPUs were used for all other training with LS-100h. 

All tasks within the SUPERB framework were executed on a single V100 GPU and followed the learning rate guidelines provided in the S3PRL toolkit \cite{yang2021superb}. To keep a comparable evaluation, we discarded the topic prediction layer when evaluating HuBERTopic with SUPERB tasks. While most tasks adhered to the default learning rates set by the SUPERB benchmark, we made specific adjustments to optimize the performance of our model, HuBERTopic. Specifically, we modified the learning rate for PR to 0.001 and for speaker identification (SID) to 0.01 to achieve enhanced performance.

\vspace{-8pt}
\section{Results}
\vspace{-5pt}
\label{sec:results}

\subsection{ASR Evaluation}
In Table~\ref{tab:asr}, we first investigated the impact of different numbers of topics $K$ on ASR performance.
LDA can be interpreted as a projection from the coordinate simplex of the pseudo-label to that of a topic and can be regarded as a kind of dimensional reduction.
Therefore, $K$ was tuned in each iteration to the extent that the number of vocabularies of pseudo tokens was not exceeded in the case of LS-100h.
The results showed that the recognition performance of the HuBERTopic outperformed that of the baseline in both training data.
This indicates that the topic classification task enhanced the ability to supplement some semantic information useful for the ASR task.
However, the rate of improvement was smaller for the 960-hour scenario.
Regarding this observation, it is possible that the appropriate value of $K$ for the auxiliary task differs depending on the training time. Additionally, there is the possibility that the benefits obtained from increasing the training data outweigh the benefits provided by the auxiliary task.
Further performance improvement for large training data is a subject for future work.

\subsection{SUPERB Benchmark}


Table~\ref{tab:superb} presents the SUPERB evaluation results across five different settings. In the evaluation conducted on the LS-100h training dataset, HuBERTopic demonstrates clear advantages over HuBERT across all tasks. This consistent outperformance suggests that HuBERTopic is highly efficient at optimizing the pre-training process. Its ability to deliver superior results even with a moderate amount of training data underlines its robustness, making it a particularly valuable asset for scenarios where data availability is limited or costly. 

When the training data volume is expanded to LS-960h, the comparative performance between HuBERT and HuBERTopic becomes less straightforward, revealing both strengths and weaknesses in different tasks. However, HuBERTopic notably excels in the PR and SD tasks. This exceptional performance in these specific tasks implies that the model benefits from additional topic guidance, likely enhancing its phonetic understanding and speaker discrimination capabilities over the original HuBERT model. 

\subsection{Topic Analysis}
\label{subsec:analysis}
The above experimental results suggest that topic labels have useful information for various downstream tasks. In his section, we analyzed what specific information the topics are tied to.
LibriSpeech corpus comprises read speech data extracted from multiple books, with each book divided into multiple chapters. Each utterance ID is linked to the speaker, gender (male/female), book, and chapter information.
We, therefore calculated the purity between the topic label and each of the attributes as follows:

\vspace{-10pt}
\begin{align}
\label{purity}
    \text{Purity}(\Omega, \Lambda) = \frac{1}{N} \sum_{k} \max_j |\omega_k \cap \lambda_j|,
\end{align}

where $\Omega=\{\omega_1, \omega_2,\cdots, \omega_N\}$ is a set of attribute labels and $\Lambda=\{\lambda_1, \lambda_2,\cdots, \lambda_N\}$ is a set of topic labels. $\max_j |\omega_k \cap \lambda_j|$ represents the number of data in attribute $k$ most frequently assigned to topic $j$.
The higher this score, the higher the relevance between the topic and the attribute label.
The results are shown in Table~\ref{tab:purity}. The data used in this experiment is for LS-100h. ``Random" refers to the result of 100 trials of purity when a topic label is randomly assigned to each utterance.
First, when $K=2$, the purity between the topic label and gender attribute is very high. In particular, when 1-iteration, i.e., the feature that generates the label, is changed from MFCC to HuBERT, the purity achieves 0.978.
This result is particularly intriguing, showing that gender classification is almost possible without supervision.
Next, for the rest of the three attributes, we calculated the purity at $K=30$, which had the best ASR performance on the dev set.
As a result, in the case of any of these three attributes, the purity is significantly higher than the random case, indicating that the topic label contains semantic information on these attributes.

\begin{table}[t]
    \centering
    \caption{Purity scores between the topic label and each of the speaker, gender, book, and chapter labels.}
    \resizebox{0.99\linewidth}{!}{


\begin{tabular}{cccccc}
\toprule
\multirow{2}{*}{\textbf{Attribute}} & \multirow{2}{*}{$K$} & \multicolumn{4}{c}{\textbf{Purity($\uparrow$)}} \\
\cline{3-6}
 & &  0-iteration & 1-iteration & \multicolumn{2}{c}{Random} \\
\cline{5-6}
 & & & & Mean & Std Dev \\
\midrule
Gender & 2 & 0.897 & 0.978 & 0.503 & 0.0012 \\
\hline
Speaker & 30 & 0.099 & 0.075 & 0.011 & 0.0002 \\
\hline
Book & 30 & 0.098 & 0.081 & 0.024 & 0.0004 \\
\hline
Chapter & 30 & 0.073 & 0.061 & 0.009 & 0.0002  \\
\bottomrule
\end{tabular}
}
    \label{tab:purity}
    \vspace{-15pt}
\end{table}

\section{CONCLUSION}
\label{sec:conclusion}
This paper proposed a new approach to enrich the semantic representation of HuBERT.
We apply a topic model to pseudo-labels to generate a topic label for each utterance. An auxiliary topic classification task is added to HuBERT by using topic labels as teachers.
This allows additional semantic information to be incorporated in a fully unsupervised manner.
Our experimental results demonstrate that our method achieves comparable or better performance than the baseline in most tasks, including the automatic speech recognition task and five out of the eight SUPERB tasks.
Moreover, we find that topic labels include various information about the utterance, such as gender, speaker, book, and chapter ID.
This highlights the effectiveness of our approach in capturing multifaceted semantic nuances, ultimately leading to improved performance in downstream tasks.

\section{Acknowledgements}
Some experiments of this work used the Bridges2 system at PSC through allocation CIS210014 from the ACCESS program, which is supported by NSF grants \#2138259, \#2138286, \#2138307, \#2137603, and \#2138296.

\vfill\pagebreak
\newpage

\section{References}
{
\printbibliography
}
\end{document}